\documentclass[seceq]{ptptex}

\usepackage{graphicx}

\title{
From QCD to nuclear matter saturation%
}

\author{
Magda \textsc{Ericson }$^{1,2,}$\footnote{ e-mail address:
magda.ericson@cern.ch}  
and
Guy \textsc{Chanfray}$^{1,}$\footnote{ e-mail address:
g.chanfray@ipnl.in2p3.fr} 
}

\inst{
$^1$ Universit\'e de Lyon, Univ.  Lyon 1, 
 CNRS/IN2P3,\\ IPN Lyon, F-69622 Villeurbanne Cedex\\
$^{2}$ Theory division, CERN, CH-12111 Geneva
}

\abst{
We discuss a relativistic chiral theory of nuclear matter with $\sigma$
and $\omega$ exchange using a formulation of the $\sigma$ model in  which all the chiral
constraints are automatically fulfilled. We establish a relation between the nuclear response
 to the scalar field and the  QCD one which includes the nucleonic parts. It allows
 a comparison between nuclear and QCD information. Going beyond the mean field approach
 we introduce the effects of the pion
loops supplemented by the short-range interaction. The corresponding Landau-Migdal
 parameters are taken from spin-isospin physics results. The parameters linked to the 
 scalar meson exchange are extracted from lattice QCD
 results. These inputs lead to a 
reasonable description of the saturation properties, illustrating the link between
 QCD and nuclear physics.  We also
derive from the corresponding equation of state the density dependence of the quark condensate
 and of the QCD susceptibilities.}

\begin{document}

\maketitle

\section{Introduction}
The subject  of this work is the interplay between nuclear physics and QCD and examples of what each field can
bring to the other. Its motivations have been twofold. The first
aim is to study QCD related quantities such as the quark condensate in the nuclear medium
or the QCD scalar susceptibility which is its derivative with respect to the quark mass, 
in a way which is fully consistent with the saturation properties of nuclear matter.
The second one is to build a relativistic theory of nuclear matter which satisfies all the chiral constraints. 
It is possible to find a common frame to reach this goal in effective theories.  
They are  built to mimick QCD at low energy, allowing the study of QCD related quantities. 
Moreover a linear realization of chiral symmetry involves  a scalar isoscalar field which can generate the nuclear
attraction, {\it i.e.}, the scalar field of the $\sigma \omega$ model of Walecka-Serot \cite {SW86}. Our final conclusion 
will be that there exists a direct and model independent link between QCD quantities and the parameters which govern 
the attraction in the $\sigma\omega$ model.  In the linear sigma model  the explicit symmetry beaking part of the Lagrangian is
${\cal L}_{\chi SB}=c\,\sigma$ where $c= f_\pi\,m_\pi^2$ and $\sigma$ is the scalar field, chiral partner of the pion. 
This quantity plays the role of the symmetry breaking Lagrangian of QCD~:
${\cal L}_{\chi SB}^{QCD}=- 2\,m_q\,\bar q\, q$.
We have then the equivalence~: 
\begin{equation}
{\bar q q(x)\over \langle\bar q
q\rangle_{vac}}= 
 {\sigma(x)\over f_\pi}.
\end{equation}  
Thus it is the expectation value of the scalar field which controls the evalution
 of the condensate~:
\begin{equation}
{ \langle\bar q q(\rho)\rangle\over \langle\bar q
q\rangle_{vac}}=
 {\langle \sigma(\rho)\rangle\over f_\pi}
\end{equation}  
 and the sigma propagator (taken at zero momentum) which controls the susceptibility~:
\begin{equation}
\chi_S=\int d^4x\,\langle \bar q
q (x) \, \bar q q (0) \rangle  ={ \langle \bar q
q\rangle _{vac }^2 \over f^2_{\pi }}\int d^4x
 \langle \sigma (x)\, \sigma (0) \rangle 
={\langle \bar q
q\rangle _{vac}^2 \over  f^2_{\pi }}  D_{\sigma }(0).
\end{equation} 
For  the nuclear binding problem we use, as in ref. \cite{CEG02},  a non linear representation, 
keeping a chiral singlet scalar field which is the radius of the chiral circle, $S$. In the vacuum $ S=f_\pi$,
and we denote $s$ the fluctuation: $S=f_\pi\,+\,s$. In the usual non linear sigma model this fluctuation
is ignored. We have shown  previously,  (ref. \cite{CEG02}) that the mean value of $s$ can be identified 
with the scalar mean field of the $\sigma \omega$ model. In this way, with the introduction of the non-linear
representation, all chiral constraints are fulfilled.  This would not be the case if the identification would be made 
instead with the non chiral invariant sigma field. 
The two fields $s$ and $\sigma$ are related through the transformation from polar to cartesian coordinates~:  
\begin{equation}
\label{Eq11}
\sigma =(f_{\pi }+s)\cos F\left( \frac{\phi }{f_{\pi }}\right) \simeq f_{\pi } +s -{{\phi}^2 \over 
2 f_{\pi}}
\end{equation}
in which terms of order $s\,{\phi}^2$ have been ignored. The effective nucleon mass is influenced by the  mean field $\bar s$~:
\begin{equation}
M^*_N(\bar s)=M_N\,\left( 1+{\bar s\over f_{\pi }}\right)\ \simeq M_N\,-\,{g_S^2  \over m^2_{\sigma}}\,\rho_S.
\label{EFFMASSN}
\end{equation}
Thus the nucleon mass and the condensate evolutions are linked but not proportional.
The term which involves the scalar density of nuclear pions ${\phi}^2$ is absent in the mass evolution.
Similarly the propagator of the  $\sigma$ field, which gives the scalar susceptibility and
the one of the $s$ field are linked but not identical, with (ref. \cite{CDEM06})~:
\begin{equation}
D_{\sigma}= D_s\,+\,{ 3  \over 2\,f^2_{\pi }}  G                                                        
\end{equation}
where G is the two-pion propagator. At low momenta the $s$ field is not coupled to two pions in contradistinction with the
$\sigma$ one which is strongly coupled (said differently the $s$ field, which is at the origin of the nuclear binding,
just decouples from low energy pions the dynamics of which is described by chiral perturbation theory).

\begin{figure}                     
\centering
\includegraphics[width=11cm,angle=0]{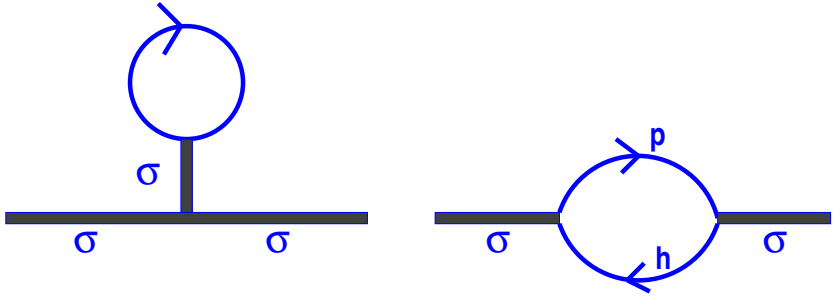} 
\caption{Sigma self-energy in the nuclear medium.}
\label{sigmaself}
\end{figure} 
In this framework a joint study of the nuclear binding and of the QCD quantities
is natural and the compatibility between the two is insured. The question is wether something more
can be learned from this common study. In a first step we ignore the pion loops. In this case the condensate evolution
is simply given by~:
\begin{equation}
{\Delta\langle\bar q q(\rho)\rangle\over \langle\bar q
q\rangle_{vac}}=
 {\bar s\over f_\pi}
\end{equation}  
where the symbol $\Delta$ for the condensate stands  for the modification with respect to the
vacuum value. Similarly the nuclar susceptibility, vacuum value subtracted, is~: 
\begin{equation}
\chi_S^A~= 
2\,{\langle\bar q
q\rangle_{vac}^2\over f^2_{\pi }}({D^*_{\sigma}} - D^0_{\sigma}).
\label{CHISA}
 \end{equation}
 Here $ D^*_{\sigma}$ is the full in-medium sigma propagator while $D_{\sigma}^0$ is the vacuum one.  Both are taken at zero momentum and they are related by~:
\begin{equation}
-(D^*_{\sigma} - D^0_{\sigma}) = 
{1\over  m^{*2}_\sigma}-{1\over  m^{2}_\sigma}\,-\,{g^2_S \over  m^{*2}_\sigma}\, \Pi_{S}(0)\,{1\over  m^{*2}_\sigma}.
\label{CHISEFF}\end{equation}
In this expression,  $g_S$ is the scalar coupling constant and $\Pi_S$ is the full RPA particle-hole polarization propagator
in the scalar-isoscalar channel. The mass $m^*_\sigma$ is the effective sigma mass which differs
 from the free one mostly by the effect of the tadpole term~:
\begin{equation}
m^{*2}_{\sigma} 
= m^{2}_{\sigma}\left(1 \,+\,{3\bar s\over f_\pi}\,+\,{3\over 2}\left({\bar s\over
f_\pi}\right)^2\right)\,\simeq  m^{2}_\sigma-{3\,g_S \over f_{\pi}}\rho_S 
\label{MSIGMA}
\end{equation}
This  represents a large decrease of the mass $\simeq30\%$ at $\rho_0$. The tadpole contribution and the effect 
of the p.h. polarization propagator are illustrated in fig. \ref{sigmaself}. The first part corresponds to the 
$\sigma$  self-energy from the following $\sigma N$ (non Born) amplitude, $T_{\sigma N}=-3g_s/f_{\pi }$,
while the second part arises from the in medium modified (Pauli blocked) $\sigma N$ Born amplitude. 
Inserting the expression  (\ref {CHISEFF}) of the sigma propagator into that, eq. (\ref {CHISA}), of the susceptibility
we obtain to lowest order in density~:
\begin{equation}
\chi _S^A\,\simeq \,2\,{\langle\bar q q\rangle_{vac}^2\over f_{\pi }^2\,m^{4}_{\sigma } }
\left[-{3\,g_S \,\rho_S \over f_{\pi}  }\,+\, g^2_S \,  \Pi_{S}(0)\right] .
\label{CHIAS}
\end{equation}
In this expression the  first term linear in the density embeds the scalar susceptibility of the individual 
nucleons while the second term reflects the effect of the nuclear excitations which  
affect the QCD susceptibility through the coupling of the quark density fluctuations to
the nucleonic ones \cite{CE03}. All in all the parenthesis on the r.h.s. of eq. (\ref{CHIAS}) is the total
nuclear response to the nuclear scalar field, ${\cal R}^A$. The proportionnality factor between  $ {\cal R}^A$ 
and $\chi^{A}_{S}$ can be expressed in terms of the scalar quark number of the nucleon arising from the $\sigma$ field~: 
\begin{equation} 
\label{QSS}
Q_S^s= {\sigma_N^s\over 2 \, m_q} = -{ \langle\bar q q\rangle_{vac}\over f_{\pi }} \int d^3 x\langle N| 
\sigma (x)|N\rangle
 = -{\langle\bar q q\rangle_{vac}\over f_{\pi }}\,{g_S\over m_{\sigma}^2}
\end{equation} 
such that~:
 \begin{equation}
\chi _S^A =2\,{( Q_S^{s})^2\over g^2_S} \, {\cal R}^A .
\label{CHIARA}
\end{equation}
It is remarkable that the full response is reflected in the QCD one (and vice versa), including the 
nucleon part \cite{INTERPLAY06}.

Coming back to  eq. (\ref{CHIAS}) the term linear in the density provides the  QCD scalar  susceptibility of 
the nucleon from the scalar field ~:
\begin{equation}
(\chi^N_S)^s \,=\,-2\,{\langle\bar q q\rangle_{vac}^2\over f_{\pi }^3}\,{3\,g_S
\over m_{\sigma}^4}. 
\label {CHISN}
 \end{equation}
This component, which to our knowledge has not been signalled before, is an example of a new information that 
the nuclear problem  can bring. The  question is wether this component  has a reality. In the model certainly, 
since it can be also be obtained as the derivative of $Q_S^s$ with respect to the quark mass. And in reality? There are
indications in favor of a negative component (beside the pionic one discussed in ref. \cite{CEG03}) in the lattice results
on the evolution of the nucleon mass with the quark mass, (equivalently the pion squared mass).
They are available only above $m_{\pi} \simeq 400\, MeV$ and in order to extract the physical nucleon mass an extrapolation has 
to be performed. Thomas et al. \cite{TGLY04} have separated out the pion which introduces a non analytical behavior in $m_q$. 
For this they evaluate the pionic part of the self-energy, $\Sigma_{\pi}$, introducing different shapes of the form factor 
at the $\pi NN $ vertex with an adjustable cut-off   parameter. They expand the rest of the nucleon mass in powers of 
$m_{\pi }^2$ as follows~:
\begin {equation}
M_N(m^{2}_{\pi}) = 
a_{0}\,+\,a_{2}\,m^{2}_{\pi}\, +\,a_{4}\,m^{4}_{\pi}\,+\,\Sigma_{\pi}(m_{\pi}).
\end{equation}
The parameters $a_i$ show little sensitivity to the shape of the form factor, with  values $a_2 \simeq 1.5\,GeV^{-1}$ 
while $a_4 \simeq- 0.5\, GeV^{-3}$, (see ref. \cite{TGLY04}). From this we can infer the non-pionic pieces of the sigma 
commutator using the Feynman-Hellman theorem~:
\begin{equation}
\sigma_N^{non-pion} = m^2_{\pi} \,{\partial M\over \partial m^2_{\pi }}
=a_2 \,m^2_{\pi}\, + \,2\,a_4 \, m^4_{\pi}\simeq 29\, MeV\,.
\label{SIGMANOPION}
\end{equation}
It is dominated by the $a_2 $ term. Its value indicates the existence of a large component  in $\sigma_N$ beside the pion 
one, that it is natural to attribute to the scalar field.  The next derivative provides the non pionic susceptibility~:
\begin{equation}
\chi_{NS}^{non- pion}~= 2{\langle\bar q q\rangle_{vac}^2 \over f^4_{\pi} }\,
{\partial~~\over \partial m^2_{\pi }}\left({\sigma_N^{non-pion} \over m^2_{\pi }}\right)=  
{\langle\bar q q\rangle_{vac}^2 \over f^4_{\pi} }\,4\,a_{4}.
\label{CHIA4}
\end{equation}
It has a negative sign, as expected from the scalar contribution. If the signs are right, are the  numerical values also 
compatible with the sigma model?  If we identify the two quantities,  $ \sigma_N^{non-pion}$ and  $ \chi_{NS}^{non- pion}$  
with the sigma model values, we first have~:
\begin{equation}
\sigma_N^{non-pion} 
\simeq a_2 \,m^2_{\pi}\, =  \sigma_N^s
=  f_{\pi }m^2_{\pi}\,{g_S\over m_{\sigma}^2}
\label{SIGMANONPION}										
\end{equation}
which leads to $m_{\sigma}=800\,MeV$  (for $g_S=M_N/f_{\pi }$). On the other hand the identification of the two expressions 
\ref{CHISN} and \ref{CHIA4} for  $ \chi_{NS}^{non- pion}$ and the  elimination of $m_{\sigma}$ using \ref{SIGMANONPION}, 
leads to the relation~:  
\begin{equation} a_4 \simeq -{3 \over 2M} a_2^2  \simeq -3.4 \,GeV^{-3}
\end{equation}
while the value found in the expansion is only   $ - 0.5\, GeV^{-3}$. The model as such fails to pass the QCD test. 
In fact this is to be expected and even gratifying because it also fails the nuclear physics test.
Indeed the effect of the tadpole term on the $ \sigma $ propagator is too large. The  softening of the sigma mass makes 
nuclear matter collapse and prevents saturation \cite {KM74}.  We have seen that the QCD susceptibility and the $\sigma N $ 
amplitude are related.  Both values are too large in magnitude in the model which  is incomplete and has to be improved.  
Indeed  an important effect is  missing, namely  the  scalar response of the nucleon,  $ \kappa_{NS}$, to the scalar 
nuclear field, which is the basis of the quark-meson coupling model, (QMC), introduced in ref.\cite{G88}.  
The crucial point is that its origin lies in confinement in such a way that  its sign is positive, 
{\it i.e,,} it opposes an increase of the scalar field. In QMC the response is calculated in a pure bag model. 
The quantity  $\kappa_{NS}$ is then related to the QCD scalar susceptibility of the bag through a 
relation similar to ours (\ref{CHIARA}) but in which only bag quantities enter. The  positive sign follows as the confined 
quarks become less relativistic with increasing quark mass and their scalar number increases. In 
order to incorporate all aspects, it will be interesting to explore the full nucleonic response to the scalar  
field in a model of the nucleon, such as the one of Shen and Toki \cite{ST00}, which incorporates both aspects, 
{\it i.e.},  where the nucleon mass originates from both the coupling to the condensate and the confinement. 
 
\begin{figure}                     
\centering
\includegraphics[width=5cm,angle=0]{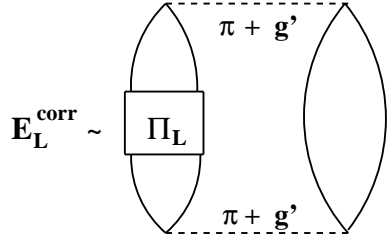} 
\caption{Schematic representation of the longitudinal spin-isospin contribution 
to the correlation energy.}
\label{encorr}
\end{figure}

\begin{figure}                     
\centering
\includegraphics[width=8cm,angle=270]{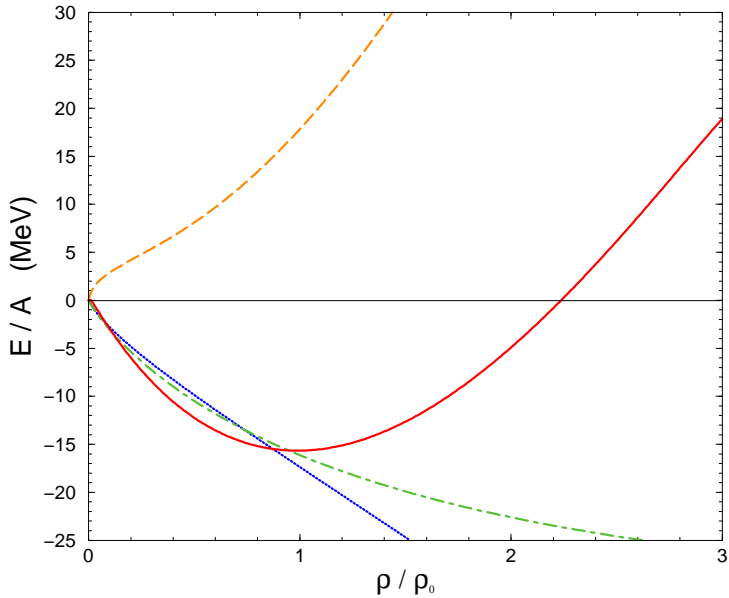} 
\caption{ Binding energy of nuclear matter with $g_\omega=8.0$, $m_\sigma=850\, MeV$  and $C=0.985$.
The full line corresponds to the full result, the dotted line represents the binding energy
without the Fock and  correlation energies and the dot-dahed line corresponds to 
the contribution of the Fock terms. The decreasing dotted line (always negative) represents the
correlation energy.}
\label{corrbind1}
\end{figure}

It is possible \cite{CE05} to improve our previous model described above with the phenomenological introduction of the scalar nucleon 
response, $\kappa_{NS}$. It modifies the nucleon mass evolution as follows,
 \begin{equation}
M^*_N(\bar s)=M_N\,\left( 1+{\bar s\over f_{\pi }}\right)\,+{1\over 2}\,\kappa_{NS}\,
\bar s^2 .\label{EFFMASSN}
\end{equation}
The sigma mass is also affected~:    
\begin{equation}
m^{*2}_{\sigma}\simeq 
m^{2}_{\sigma}- ( {3\,g_S \over f_{\pi}} -  \, \kappa_{NS})\,\rho_S ,
\label{MSIGMA}
\end{equation}
and the nucleonic QCD susceptibility as well, in such a way that the relation between $a_4$ and $a_2$ becomes~:
\begin{equation}
a_4= -{ a_2^2 \over 2\,M}( 3\,-\,2\,C).
\end{equation}
where $C$ is the dimensionless parameter $ C= \left({f^2_{\pi}/  2\,M} \right) \kappa_{N S} $. Numerically 
$a_4=- 0.5 \,GeV^{-3}$ gives $ C =+1.25$. A large cancellation of the tadpole effect is indeed required by the 
lattice expansion, total cancellation occuring for $C=1.5$.
Our approach has then consisted in going from QCD to nuclear physics and use the lattice expansion 
to fix the parameters of the $\sigma\omega$ model \cite{CE06}. The question is wether this procedure makes sense and 
wether it leads to a possible description of the nuclear binding.  We have explored this, first in the mean field approach, 
but also with the introduction of the pion loops  \cite{CE06}, as is described below. In the latter case the 
condensate acquires a new component linked to the nuclear pion scalar density, $\langle \phi ^2  \rangle $. 
However the nucleon mass   is  blind to this component \cite{CEG02}. This is also true for the sigma  mass and the 
eq. (\ref{EFFMASSN}) and (\ref{MSIGMA}) remain valid.  Pion loops nevertheless affect  the energy. They  do not enter at 
the mean field level but  contribute through the Fock term and through the correlation energy schematically depicted 
in fig. \ref{encorr}. It includes iterated pion exchange and also the part of the NN potential from the two-pion exchange 
with $\Delta$ excitation.  In the correlation term we have introduced beside pion exchange the short-range components 
{\it via} the Landau-Migdal parameters, $g'_{NN},\, g'_{N\Delta},\, g'_{\Delta\Delta}$. We have taken their values from  
a systematic survey of the data on spin-isospin physics  by Ichimura {\it et al.} \cite{ISW06}.  For the parameters $g' $ 
they indicate large deviations from universality with~: $g'_{NN}=0.7,\,  g'_{N\Delta }=0.3,\,  g'_{\Delta \Delta }=0.5 $. 
Beside pion exchange  we have also introduced the transverse channel, dominated by $\rho $ exchange together with the 
short-range component. The other parameters of the model are chosen as follows. The form factor at the  $\pi NN $ 
vertex is taken as a dipole  with a cutoff parameter $\Lambda = 7\,m_{\pi }$, which is guided by the following 
considerations~: it leads to a pionic contribution to $\sigma _N$ of $22\,MeV$.  Adding  the lattice value of the non-pionic part, 
$\sigma ^{non-pion} = 29\,MeV$, the sum takes the  value  $\sigma_N= 51\,MeV$, in the well accepted range. The scalar 
coupling constant is the one of the model $g_S= M/f_{\pi} =10 $ and for the sigma mass we have followed the lattice 
indications, allowing a small readjustment around the lattice value, which is $m_{\sigma}=800\, MeV$. We have found a better 
fit with $m_{\sigma }=850\,MeV$ which corresponds to  $\sigma^{non-pion} = 26\,MeV$. The omega mass is known and the 
$\omega NN$ coupling constant totally free. For the nucleon scalar response we have followed the indications of the 
lattice data but not strictly in view of the uncertainties attached to the higher derivatives. The value which fits 
the saturation properties is found to be $C\simeq 1$, not far from the lattice value $C= 1.25$. With these inputs 
we have obtained a satisfactory description of the nuclear binding.  The binding energy per particle is shown  in 
fig. \ref{corrbind1}, with its different components. We like to comment on the correlation energy which has a value 
of $-\,17.4\,MeV$ . The longitudinal channel is strongly suppressed by the short-range component, (with less cancellation 
in its  $N\Delta$ component). This leaves the transverse part as the dominant contribution ($-\,9.5\,MeV$  
{\it versus} $-\, 7.9\,MeV$  for the longitudinal one).
 \begin{figure}                  
\centering
\includegraphics[width=8cm,angle=270]{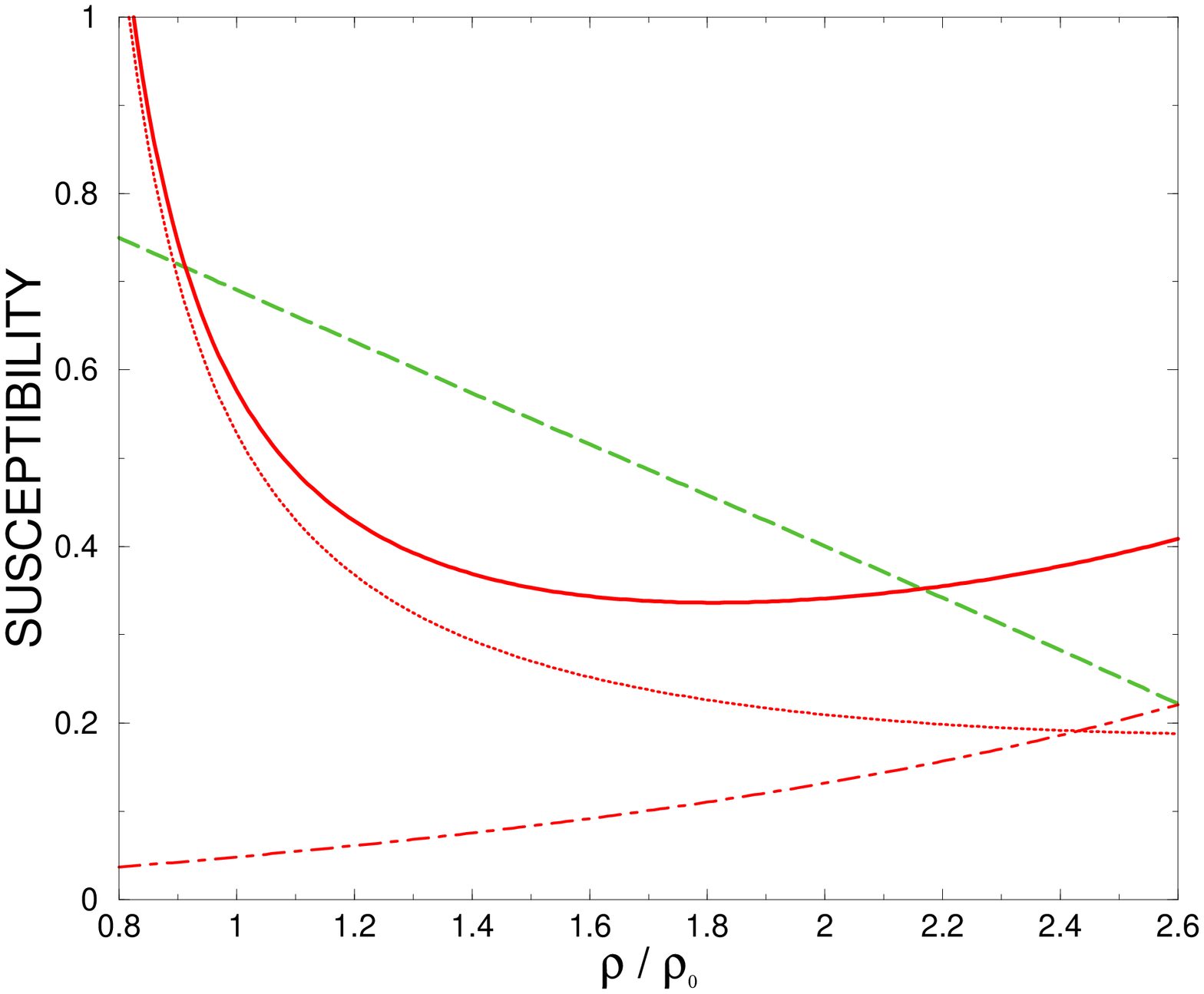} 
\caption {Density evolution of the QCD susceptibilities (normalized to the
 vacuum value of the pseudoscalar one)  with $g_\omega=8.$, $m_\sigma=850\, MeV$ and $C=0.985$.
Dashed curve: pseudoscalar susceptibility. Full curve: Scalar susceptibility. Dotted curve:
nuclear contribution to the scalar susceptibility. Dot-dashed curve: pion loop contribution to the
scalar susceptibility.}
\label{corrsusc1}
\end{figure}
From the equation of state and using  the Feynman-Hellmann theorem we have deduced the quark condensate. The next derivative 
with respect to $m_q$ provides the scalar susceptibility.  The two QCD susceptibilities, scalar and pseudoscalar  are 
depicted in fig \ref{corrsusc1} . The second one follows the evolution of the condensate \cite{CE03} and it  is remarkably 
linear in the density in spite of the interaction. The scalar one shows a strong increase with respect to the free value, 
surpassing even the pseudoscalar one around $2\rho_0$. This behavior is largely due to the mixing with the nuclear excitations.
\\

In summary we have studied a chiral relativistic theory of nuclear matter based on the sigma model and its interplay with 
QCD. We have shown that the softening of the sigma mass  arising from the tadpole term, (from $3\,\sigma$ interaction) has 
a counterpart in QCD in the form of a negative contribution to the nucleonic QCD scalar susceptibility from the scalar field.  
We have investigated the presence of this component in the lattice expansion of the nucleon mass with respect to the 
quark mass.  There is indeed an indication in favor of a negative component but its magnitude is much too small. This is 
in fact totally consistent with  nuclear physics. The only effect of the tadpole with the softening of the $\sigma$ mass 
prevents saturation. In both cases a canceling effect must occur. For this we have introduced, as in QMC, the scalar 
response of the nucleon which is a reflect of the QCD scalar susceptibility of the nucleon which is due to confinement. 
Our approach has then be to utilize the QCD information  on the nucleon mass evolution with the quark mass 
to fix or constrain the parameters of the $\sigma\omega$ model. This, together with the information from spin-isospin 
physics, has allowed a successfull description of the binding properties of nuclear matter.   
\section*{Acknowledgments} 
One of the authors (M. 
E.) thanks the Yukawa Institute for Theoretical
Physics at Kyoto University, for the hospitality  during the YKIS2006 on
"New Frontiers on QCD".


%


\end{document}